\documentstyle[12pt]{article} \begin{document} \thispagestyle{empty}
\begin{center} \LARGE \tt \bf {Decoupling between torsion and magnetic fields in bouncing cosmology and galactic dynamo seeds} \end{center}

\vspace{3.5cm}

\begin{center} {\large L.C. Garcia de Andrade \footnote{Departamento de
F\'{\i}sica Te\'{o}rica - IF - UERJ - Rua S\~{a}o Francisco Xavier 524,
Rio de Janeiro, RJ, Maracan\~{a},
CEP:20550.e-mail:garcia@dft.if.uerj.br} and} \end{center}
\begin{center} {\large A. Ferrandez\footnote{Departamento de
Matematicas-Universidad de Murcia-Campus de Espinardo, 30100,
Murcia, Spain. e-mail:aferr@um.es}} \end{center}

\begin{abstract} Recently Salim et al [JCAP (2007)], have shown that galactic dynamo seeds can be possibly attainable in bouncing cosmological models with QED Lagrangeans. In this paper we generalise their result by include torsion of spacetime in bouncing cosmology. It is shown that by considering a semi-minimal photon-torsion coupling and a Lagrangean of the type $RF^{2}$ it is possible to find a fast decoupling between magnetic and torsion fields in the contracting phases of the universe. Besides torsion field decays as $K\sim{a^{\frac{2}{3}}}$ while the magnetic field grows as $B\sim{a^{-5.5}}$ thus explaining the fast decoupling between the two fields. It is expected that at some point of the contracting phase the amplification of the magnetic field may give rise to a enough strong magnetic field to seed a galactic dynamo.

\end{abstract} Key-words: torsion theories,
bouncing cosmology, astro-particle physics. \newpage

\section{Introduction} K Enqvist et al \cite{1} place limits on neutrino masses from galactic dynamo
mechanism. Since neutrino masses are important in extending the standard
model of particle physics, it seems worth to investigate the relation between
Lorentz Violation (LV) and galactic dynamos in torsion fields for example
\cite{2,3}. Therefore knowledge of the dynamics between torsion and cosmic
magnetic fields may reveal if dynamo mechanism is a powerful mechanism to
feed the galactic magnetic fields of nano-Gauss observed in nature. In this
letter one shows that by using a scalar electrodynamics in the context
of quantum electrodynamics (QED) \cite{4} it is possible to show that magnetic
field decays when torsion is fast amplified. It is shown that torsion needed
to seed galactic dynamo is of the other of $10^{-18}cm^{-1}$ which can be found in
nature and is even weaker than the value estimated in the Early Universe.
In previous work \cite{3} one notice that semi-minimal coupling has been used
on a Lagrangean of the type $\frac{1}{4}R_{ijkl}F^{ij}F^{kl}$, [i,j=0,1,2,3]. This
has provided further constraints on torsion up to $10^{-31}GeV$. Here though
semi-minimal coupling is preserved, we shall use, another gravitational sector
in the Lagrangean given by the coupling $RF^{ij}F_{ij}$ as used in the paper by
Mazziteli et al \cite{5}. The term $R_{ijkl}F^{ij}F^{kl}$ displays the same symmetries of LV
term, where the Riemann-Cartan curvature tensor, including torsion terms
plays the role of the Higgs sector constants $k_{ijkl}$. Here we will not address the
problem of the LV since the Lagrangean term used does not favor that. In
this paper, we show that the use of this photon sector coupled semi-minimally
with torsion mode, in scales of 10 kpc would require a not very strong torsion field that might exist
in nature, so it seems that the only conclusion is that this necessary imply
that galactic magnetic fields can be seeded by a such torsion models also
in the Mazziteli et al scalar electrodynamics is concerned with semi-minimal
coupling of course. Some physicists \cite{6} argued that torsion is very weak to
have time to seed magnetic fields but actually here from Mazziteli et al scalar
QED we have shown that the torsion field may grow exponentially in regions
of weak primordial magnetic field which is not strong enough to seed galactic dynamos. The paper is organised as follows: In section II we
review Mazziteli et al scalar electrodynamics. In section III we apply the
semi-minimal coupling to their equations to introduce the torsion field which grows in the contracting phase of the universe as in a bouncing model \cite{7}, presenting a decoupling between torsion to magnetic fields. Section IV contains
conclusions and discussions.
\section{Flat semi-minimal torsion-photon coupling
of $RF^{2}$ Lagrangean}
Though torsion effects are highly suppressed in comparison with curvature
ones of Einstein gravity sector, we do not consider here Minkowski space since as can be easily shown here from the field equations that torsion vanishes in Minkowski space. Mazziteli et
al Lagrangean \cite{5} is
\begin{equation}
S=\frac{1}{m^2}\int{d^{4}x(-g)^{\frac{1}{2}}(-\frac{1}{4}F^{2}+(m^{2}+{\epsilon}R){\phi}\bar{\phi}-D_{j}
{\phi}D^{j}\bar{\phi})}
\label{1}
\end{equation}
where $D_{i}={\partial}_{i}-ieA_{i}$ is the covariant derivative for the scalar fields. Spedalieri
et al \cite{5} have computed an effective Lagrangean for the e.m field by integrating the quantum scalar field. Via dimensional regularisation they obtain the
effective  Lagrangean \cite{7} \begin{equation}
{\cal{L}}_{eff}=-\frac{1}{4}F^{2}+\frac{1}{2}\frac{1}{4{\pi}^{\frac{d}{2}}}(\frac{m}{\mu})^{d-4}\sum{a_{j}(x)
m^{4-2j}
{\Gamma}(j-\frac{d}{2})}\label{2}
\end{equation}
The first Schwinger-De Witt (SDW) coefficientsSpedelieri et al work
\begin{equation}
a_{0}=1
\label{3}
\end{equation}
\begin{equation}
a_{1}=-(\epsilon-\frac{1}{6})R
\label{4}
\end{equation}
\begin{equation}
a_{2}=\frac{1}{180}(R_{ijkl}R^{ijkl}-R_{ij}R^{ij})+\frac{1}{2}({\epsilon}-\frac{1}{6})^{2}R^{2}+\frac{1}{6}({\epsilon}-\frac{1}{5}){\Box}R-\frac{e^{2}}{12}F^{2}
\label{5}
\end{equation}
\begin{equation}
a_{3}=...+\frac{e^{2}}{60}R_{ijkl}F^{ij}F^{kl}-\frac{e^{2}}{90}R_{ij}F^{ik}F^{kl}+(\frac{1}{6}-{\epsilon})RF^{2}+...
\label{6}
\end{equation}
where we have omitted the Maxwell terms that will not be of our interest in
the sequence of the paper. Here we note that due to the use of semi-minimal
coupling where torsion, which is our only gravitational field, appears only
in $a_{2}$ as first term, since in the semi-minimal coupling torsion does not ap-
pears in the covariant derivative and consequently not in the electromagnetic
field. Actually following this reasoning torsion appears only in the curvatures
appears for the first time in $a_{2}$. Following them I shall consider the follow-
ing effective Lagrangean in Riemann-Cartan spacetime, through the minimal
coupling as
\begin{equation}
{\cal{L}}_{eff} = -\frac{1}{4}F^{2}(1 + \frac{b}{m^{2}}R)\label{7}
\end{equation}
where we have taken n = 1 such as in Widrow and Turner \cite{8}. From this
effective Lagrangean we obtain the field equations for the Friedmann spatially
flat metric
\begin{equation}
ds^{2} = a^{2}(-d{\eta}^{2} + dx^{2}) \label{8}
\end{equation}
as
\begin{equation}
{\partial}^{i}(F_{ij}(1 + \frac{bR}{m^{2}}))=0
\label{9}
\end{equation}
From these equations one may obtain with appropriated approximations
\begin{equation}
[\ddot{A}_{k} + k^{2}A_{k}](1 + \frac{bR}{m^{2}})+\frac{b\dot{R}}{m^{2}R}{\dot{A}}_{k}=0
\label{10}
\end{equation}
these equations one may yet approximate for high coherence scales where
$k^{2} << 1$. We also addopt here the fact that in Riemannian case in inflation-
ary epoch $R >>> m^{2}$ so this would reduce the last equation to
\begin{equation}
[\ddot{A}_{k} + \frac{\dot{R}}{R}{\dot{A}}_{k}]=0
\label{11}
\end{equation}
where R is the Ricci scalar. This shows that although there is no inflation
here we consider that torsion has a similar behaviour so actually $\dot{K}
>>> m^{2}$.
\section{Galactic dynamo seeds in $R F^{2}$ semi-minimal coupling}
In this section we shall solve equation (11) in the case of curved spacetime with and performing the semi-minimal coupling where the
Ricci scalar is approximated taken as $2\dot{K}$, where K is the time component $K^{0}$,
which to simplify matters is the only homogeneous component of contortion,
an algebraic combination of torsion. Here we addopt linearisation of the
Ricci-Cartan scalar \cite{7}
\begin{equation}
R = g_{ij}R^{ij} = R^{*} + 2{\nabla}_{i}K^{i}-K^{2} \label{12}
\end{equation}
where $K^{j}={K^{rj}}_{r}$, represents the trace of torsion tensor , $R^{*}$
is the Riemannian Ricci scalar that here shall be taken as constant like in de Sitter or Einstein space, to simplify computations. Minkowski space where it vanishes can be also addressed. Let us now perform the variation of the Lagrangean density ${\sqrt{g}}{\cal{L}}$ with respect to the scale cosmological factor $a$, and contortion K, to complete the system of Einstein-Cartan-Maxwell equations of course with propagating torsion. This can be done easily by computing the Euler Lagrange equations
\begin{equation}
\frac{d}{dt}\frac{{\partial}\sqrt{g}\cal{L}}{{\partial}\dot{a}}-\frac{{\partial}\sqrt{g}\cal{L}}{{\partial}a}=0
\label{13}
\end{equation}
\begin{equation}
\frac{d}{dt}\frac{{\partial}\sqrt{g}\cal{L}}{{\partial}\dot{K}}-\frac{{\partial}\sqrt{g}\cal{L}}{{\partial}K}=0
\label{14}
\end{equation}
Let us start from the last equation to determine K in terms of the scale factor a. This yields
\begin{equation}
K=-\frac{3\dot{a}}{a}
\label{15}
\end{equation}
Before applying this result to the expression for the Ricci-Cartan scalar, let us express this scalar in terms of the scalar a and torsion trace K. This yields the following expression
\begin{equation}
R = g_{ij}R^{ij} = R^{*} + 2{\dot{K}}^{i}-K^{2}+{\partial}_{t}ln{\sqrt{g}}K
\label{16}
\end{equation}
or
\begin{equation}
R = g_{ij}R^{ij} = -6[\frac{\ddot{a}}{a}+(\frac{\dot{a}}{a})^{2}] + 2{\dot{K}}^{i}-K^{2}+({\partial}_{t}ln{a^{3}})K
\label{17}
\end{equation}
which yields
\begin{equation}
\dot{R}=\dot{R^{*}}+\ddot{K}+(\frac{\dot{a}}{a}+2K)\dot{K}+3[\frac{\ddot{a}}{a}-\frac{{\dot{a}}^{2}}{a^{2}}]K
\label{18}
\end{equation}
The expression for $\ddot{K}$ is
\begin{equation}
\ddot{K}=-3[\frac{\ddot{\dot{a}}}{a}-3\frac{\ddot{a}\dot{a}}{a^{2}}+(\frac{\dot{a}}{a})^{3}]
\label{19}
\end{equation}
The expression for Ricci-Cartan scalar Lagrangean $\sqrt{g}R$ is
\begin{equation}
a^{3}R=-3[3\ddot{a}{a}^{2}+7{\ddot{a}}^{2}a]
\label{20}
\end{equation}
Substitution of this expression into the Euler-Lagrange equation above one has
\begin{equation}
\ddot{\dot{a}}{a}-4\ddot{a}\dot{a}=0
\label{21}
\end{equation}
By making use of the ansatz $a\sim{t^{n}}$ where n is a real number, one obtains the following algebraic equation
\begin{equation}
n(n-2)-4n^{2}=0
\label{22}
\end{equation}
which yields immeadiatly $n=\frac{2}{3}$, and $a\sim{t^{-1}}$, which represents a contracting phase of the cosmological model with torsion. Therefore from the above expression for K one obtains $K\sim{a^{\frac{2}{3}}}$. On the other hand the magnetic field undergoes a dynamo like phase undergoing an amplification according to the law
$B\sim{a^{-5.5}}$. This can be easily seen by computing the ratio $\frac{\dot{R}}{R}$ as
\begin{equation}
\frac{\dot{R}}{R}=\frac{[3\ddot{K}-\frac{2}{3}(K^{2})^{.}]}{[3\dot{K}-\frac{2}{3}K^{2}]}
\label{23}
\end{equation}
Since the torsion is a very weak field this can be approximated to
\begin{equation}
\frac{\dot{R}}{R}\approx{\frac{\ddot{K}}{\dot{K}}}
\label{24}
\end{equation}
Substitution into the Maxwell like equation above and solving it, and taking into account the expression
\begin{equation}
B=ikA
\label{25}
\end{equation}
where B is the magnetic field and k is the wave number which is given by the inverse of the coherent scale, one obtains the above value of $B\sim{a^{-5.5}}$.
Therefore one may conclude that the torsion decays while the magnetic field grows in the contracting phase of the universe \cite{8} exactly like in the general relativistic version investigated by Salim et al \cite{9}.
\section{Discussion and conclusions}
In this paper we show that a semi-minimal torsion coupling in Riemann-Cartan spacetime can be used to generalize previous papers on QED cosmology where a Minkowski space with torsion has been used. AsLagrangeans can be used to determine the torsion which can be used to seed galactic dynamos. The motivation from this
study came from some work by Campanelli et al \cite{2} where they investigate
similar subjects in the general relativistic backgrounds of Riemannian geometry and by the work of Salim et al \cite{9} on the amplification of the magnetic field in bouncing cosmological models. As in their case we think some improvements can be made in obtained galactic dynamo seeds if one uses the matter content of the universe instead of the vacuum QED in curved spacetimes with torsion used here. This may appear elsewhere.
\section{Acknowledgements}
 We would like to express our gratitude to V. Lemes, Ariel Zithnitsky, A Maroto and D Sokoloff for
helpful discussions on the subject of this paper. Special thanks n anonymous referee for his extremely kind suggestions which allowed us to making considerable and important improvements on the first draft of this letter. One of us (GdA) would like to thank financial supports from CNPq. and University of State of Rio de Janeiro (UERJ).

\end{document}